\documentclass{JHEP3}

\usepackage{amsmath}
\usepackage{amsfonts}
\usepackage{cite}
\usepackage{subfigure,epsfig}
\usepackage{psfrag}
\usepackage{appendix}
\usepackage{booktabs}

\def\e{\epsilon}

\title{Integrated dipoles with MadDipole in the MadGraph framework}

\author{R.\ Frederix$^a $ T.\ Gehrmann$^a$,
N.\ Greiner$^{a,b}$\\
$^a$ Institut f\"ur Theoretische Physik, Universit\"at Z\"urich,\\
Winterthurerstrasse 190, 8057 Z\"urich, Switzerland\\
$^b$
Department of Physics, University of Illinois at Urbana-Champaign,\\
1110 West Green Street, Urbana, IL 61801, USA}

\preprint{ZU-TH 06/10}

\abstract{ Heading towards a full automation of
  next-to-leading order (NLO) QCD corrections, one important
  ingredient is the analytical integration over the one-particle phase
  space of the unresolved particle that is necessary when adding the
  subtraction terms to the virtual corrections. We present the
  implementation of these integrated dipoles in the MadGraph
  framework. The result is a package that allows an automated
  calculation for the NLO real emission parts of an arbitrary process.
}

\begin{document}
\newcommand{\bra}[1]{\langle#1|}
\newcommand{\ket}[1]{|#1\rangle}
\newcommand{\nn}{\nonumber}
\newcommand{\zi}{\tilde z_i}
\newcommand{\zj}{\tilde z_j}
\newcommand{\sijk}{s_{ij,k}}
\newcommand{\yijk}{y_{ij,k}}
\newcommand{\vijk}{v_{ij,k}}
\newcommand{\viji}{v_{ij,i}}                                                    
\newcommand{\tvijk}{\tilde v_{ij,k}}
\def\aand{\!\!\!\!\!\!\!\!&&}
\newcommand{\eps}{\epsilon}
\newcommand{\alps}{\alpha_{\mathrm{s}}}
\newcommand{\CF}{C_{\mathrm{F}}}
\newcommand{\CA}{C_{\mathrm{A}}}
\newcommand{\TR}{T_{\mathrm{R}}}                                                
\newcommand{\Nc}{N_{\mathrm{c}}}       
\newcommand{\bV}       {{\bf V}}
\newcommand{\cV}       {{\cal V}}
\newcommand{\cK}       {{\cal K}}
\newcommand{\cT}       {{\cal T}}
\newcommand{\rd}{{\mathrm{d}}}
\newcommand{\tpij}{\widetilde p_{ij}}
\newcommand{\tpk}{\widetilde p_k}
\newcommand{\coll}{{\mathrm{coll}}}
\newcommand{\eik}{{\mathrm{eik}}}
\newcommand{\xija}{x_{ij,a}}
\newcommand\Oe[1]      {\ensuremath{\mathrm O(\ep^{#1})}}
\def\ep{\epsilon}
\newcommand\as{\alpha_{\mathrm{s}}}
\def\bom#1{{\mbox{\boldmath $#1$}}}
\def\KFS#1{K^{{#1}}_{\scriptscriptstyle\rm F\!.S\!.}}
\def\HFS#1{H_{{#1}}^{\scriptscriptstyle\rm F\!.S\!.}}
\def\Kab{\KFS{ab}}
\section{Introduction}

Multi-particle final states are the basis of many physics studies at
the CERN LHC. In searching for physics beyond the standard model, one
is aiming to identify new particles from their decay products, which
could often result from decay chains. Likewise, accompanying final
state particles may help to improve the ratio of signal to background
processes, as done for example in the Higgs boson search through the
vector boson fusion channel.  Meaningful searches for these signatures
require not only a very good anticipation of the expected signal, but
also of all standard model background processes which could result in
identical final state signatures. From the theoretical point of view,
high precision implies that one has to go beyond the leading order in
perturbation theory to be able to keep up with the precision of the
measurements.

For leading order processes there have been many developments
concerning event generation and simulation tools in the last two
decades such as MadGraph/MadEvent
\cite{Stelzer:1994ta,Maltoni:2002qb,Alwall:2007st} CompHEP/CalcHEP
\cite{Boos:2004kh}/\cite{Pukhov:2004ca}, SHERPA
\cite{Gleisberg:2003xi,Gleisberg:2008ta} and WHIZARD
\cite{Kilian:2007gr} and also programs using different approaches such
as ALPGEN \cite{Mangano:2002ea} and HELAC \cite{Cafarella:2007pc}.
All these programs are multi-purpose event generator tools, which are
able to compute any process (up to technical restrictions in the
multiplicity) within the standard model, or within alternative
theories specified by their interaction Lagrangian or Feynman
rules. They usually provide event information which can be passed into
parton shower, hadronization and/or detector simulation through
standard interfaces.

Next-to-leading order (NLO) calculations are at present performed on a
process-by-process basis. The widely-used programs MCFM
\cite{Campbell:1999ah,Campbell:2002tg}, NLOJET++ \cite{Nagy:2003tz},
MC@NLO
\cite{Frixione:2002ik,Frixione:2003ei,Frixione:2005vw,Torrielli:2010aw}
and programs based on the POWHEG method
\cite{Nason:2004rx,Nason:2006hfa,LatundeDada:2006gx,Frixione:2007nw,Alioli:2008gx,Hamilton:2008pd}
collect a variety of different processes in a standardized framework,
the latter two methods also match the NLO calculation onto a parton
shower. The POWHEG box~\cite{Alioli:2010xd} provides a toolkit for
adapting further NLO calculations to match onto parton showers.

The NLO QCD corrections to a given process with a $n$-parton final
state receive two types of contributions: the one-loop virtual
correction to the $(2\to n)$-parton scattering process, and the real
emission correction from all possible $(2\to n+1)$-parton scattering
processes.  For the numerical evaluation, one has to be able to
compute both types of contributions separately.
 
The computation of one-loop corrections to multi-particle scattering
amplitudes was performed on a case-by-case basis up to now, the
calculational complexity increased considerably with increasing number
of external partons. Since only a limited number of one-loop integrals
can appear in the final result~\cite{Melrose:1965kb,Passarino:1978jh,Bern:1993kr}, the
calculation of one-loop corrections can be reformulated as the
determination of the coefficients of these basis integrals, plus
potential rational terms.  Based on this observation, a variety of
methods for the systematic determination of the one-loop integral
coefficients and the rational terms have been formulated, and first
fully automated programs for the calculation of one-loop multi-parton
amplitudes are becoming available with the packages
CutTools~\cite{Ossola:2007ax,Ossola:2008xq},
BlackHat~\cite{Berger:2008sj}, Rocket~\cite{Giele:2008bc} and
GOLEM~\cite{Binoth:2008uq}, as well as independent
libraries~\cite{Denner:2005nn}.

These recent technical advances have allowed the calculation of NLO
corrections to several $2\to 4$ processes, which is the current
frontier in complexity. The first result in this class of processes
were the electroweak corrections to four-fermion production in
electron-positron annihilation~\cite{Denner:2005fg}, based
on~\cite{Denner:2005nn}. Most recently NLO QCD corrections were
obtained for genuine $2\to 4$ processes at hadron colliders: $W+3$~jet
production~\cite{Berger:2009zg,Berger:2009ep,Ellis:2009zw,KeithEllis:2009bu},
$Z+3$~jet production~\cite{Berger:2010vm},
$pp\to t\bar t b\bar
b$~\cite{Bredenstein:2009aj,Bredenstein:2010rs,Bevilacqua:2009zn} and
$pp\to t\bar t jj$~\cite{Bevilacqua:2010ve}, as well as for the
quark-antiquark contribution to $pp\to b\bar b b\bar
b$~\cite{Binoth:2009rv}.

The real emission corrections contain soft and collinear
singularities, which become explicit only after integration over the
appropriate real radiation phase space yielding a hard $n$-parton
final state. They are canceled by the singularities from the virtual
one-loop contributions, thus yielding a finite NLO correction.  To
systematically extract the real radiation singularities from arbitrary
processes, a variety of methods, based either on phase-space
slicing~\cite{Giele:1991vf} or on the introduction of
process-independent subtraction terms~\cite{Kunszt:1992tn} have been
proposed. Several different algorithms to derive subtraction terms are
available: residue (or FKS) subtraction~\cite{Frixione:1995ms} and
variants thereof~\cite{Somogyi:2009ri}, dipole
subtraction~\cite{Catani:1996vz,Catani:2002hc} and antenna
subtraction~\cite{Kosower:1997zr,Campbell:1998nn,GehrmannDeRidder:2005cm,Daleo:2006xa}.

Especially the dipole subtraction formalism, which provides local
subtraction terms for all possible initial and final state
configurations~\cite{Catani:1996vz} and allows to account for
radiation off massive partons~\cite{Catani:2002hc}, is used very
widely in NLO calculations.  The generation of dipole terms for
subtracting the singular behaviour from the real radiation
subprocesses has been automated in various event generators: in the
SHERPA framework \cite{Gleisberg:2007md}, the TeVJet framework
\cite{Seymour:2008mu}, the HELAC framework~\cite{Czakon:2009ss} and in
the form of independent libraries~\cite{Hasegawa:2009tx} interfaced to
MadGraph. The MadDipole package~\cite{Frederix:2008hu} provides an
implementation within MadGraph. An implementation of the residue
subtraction method is also available within
MadGraph~\cite{Frederix:2009yq}.

For a full NLO calculation, the dipole terms have to be integrated
over the dipole phase space, and added with the virtual corrections to
obtain the cancellation of infrared singularities. So far, only one of
the implementations~\cite{Czakon:2009ss} provides these integrated
dipole terms including all masses and possible phase-space
restrictions, and constructs the corresponding integrated subtraction
terms. It is the purpose of this paper to implement the generation of
the integrated subtraction terms into MadDipole, which will
consequently allow to carry out the full dipole subtraction within the
MadGraph/MadEvent framework. The output results are {\tt Fortran}
subroutines which return the squared amplitude for all possible
unintegrated and integrated dipole configurations in the usual
MadGraph style.

With a complete treatment of the real NLO radiation, MadDipole is a
crucial building block of automated NLO calculations. This automation
is a very high priority for LHC predictions~\cite{Binoth:2010ra}, and
can likely be accomplished only in a collaborative effort, with
different groups supplying different pieces, linked through standard
interfaces~\cite{Binoth:2010xt}.

The paper is structured as follows: in Section~\ref{sec:struct}, we
briefly review the structure of unintegrated and integrated dipole
terms, Section~\ref{sec:exp} discusses the expansion of the integrated
dipoles, and normalization conventions used in this. The MadDipole
implementation of the integrated dipoles is described in
Section~\ref{sec:prog}, with functionality checks documented in
Section~\ref{sec:checks}. Finally, we conclude with
Section~\ref{sec:conc}.

\section{Structure of NLO dipole subtraction terms}
\label{sec:struct}
The fundamental building blocks of the subtraction terms in the dipole
formalism~\cite{Catani:1996vz,Catani:2002hc} are dipole splitting
functions ${\bf V}_{ij,k}$, which involve only three partons: emitter
$i$, unresolved parton $j$, spectator $k$. A dipole splitting function
accounts for the collinear limit of $j$ with $i$, and for part of the
soft limit of $j$ in between $i$ and $k$.  The dipole factors, which
constitute the subtraction terms, are obtained by multiplication with
reduced matrix elements, where partons $i$, $j$ and $k$ are replaced
by recombined pseudo-partons $\widetilde{ij}$, $\widetilde{k}$.  The
full soft behavior is recovered after summing all dipole factors.

Throughout the whole paper we are using the notation introduced in
Refs.~\cite{Catani:1996vz} and \cite{Catani:2002hc}. The dipole
factors are subtracted from the real radiation contribution at
NLO. They subtract the singular contributions where one parton from
the NLO real radiation contribution becomes soft and/or collinear,
such that the phase-space integral of this contribution can be
preformed numerically, including arbitrary kinematic restrictions on
the final-state phase space.

The dipole factors are integrated analytically over the dipole phase
space (which fully includes the infrared singular regions), such that
the integrated dipole factors have the same kinematic structure as the
virtual one-loop NLO corrections and the collinear counterterms from
mass factorization.  These integrated dipole contributions can then be
added to the virtual and mass-factorization corrections, thereby
accomplishing the cancellation of infrared singularities.

Several algorithms to automatically generate the unintegrated dipole
terms for arbitrary processes have been devised and implemented in
various matrix element generator frameworks. They are available for
SHERPA~\cite{Gleisberg:2007md}, HELAC~\cite{Czakon:2009ss}, for
MadGraph~\cite{Frederix:2008hu} as well as a implementation of
stand-alone routines~\cite{Seymour:2008mu,Hasegawa:2009tx}. Also
implementations of integrated dipoles are available in the same
codes. However, only the implementation based on the HELAC framework
has the full set of integrated dipoles for arbitrarly masses and
phase-space restrictions~\cite{Czakon:2009ss}. It was
used in the context of the calculations of NLO corrections to $pp\to
t\bar t b\bar b$~\cite{Bevilacqua:2009zn} and $pp\to t\bar t
jj$~\cite{Bevilacqua:2010ve}. In this work, we document our
implementation of the integrated dipoles in MadDipole, which is a
package in the MadGraph framework. With this extension, MadDipole
computes the full NLO dipole subtraction of real radiation and
performs the infrared cancellations in a fully automated manner
for color- and helicity-summed matrix elements squared. Our
implementation was already used in the computation of the NLO
corrections to $pp\to b\bar b b \bar b$ in the quark-initiated
channel~\cite{Binoth:2009rv}.

\subsection{Phase space restriction: $\alpha$-parameter}
\label{sec:alpha}
The calculation of the subtraction terms is only necessary in the
vicinity of a soft and/or collinear limit. Away from these limits the
amplitude is finite and there is in principle no need to calculate the
computationally heavy subtraction terms.  The distinction between
regions near to a singularity and regions without need for a
subtraction can be parametrized by a parameter
usually labelled $\alpha$ with $\alpha \in [0,1]$, which was
introduced in Ref.~\cite{Nagy:1998bb} for processes involving partons only
in the final state. The case with incoming hadrons, \textit{i.e.}, with
partons in the initial state, is described in Ref.~\cite{Nagy:2003tz}.

Using the notation of Ref.~\cite{Nagy:2003tz}, the contribution from
the subtraction term to the differential cross section 
in the real radiation channel can be written
as
\begin{eqnarray} \nn
\lefteqn{ d\sigma_{ab}^A = \sum_{\{n+1\}} 
d\Gamma^{(n+1)}(p_a,p_b,p_1,...,p_n+1)
  \frac{1}{S_{\{n+1\}}}}\\ \nn
  &&\quad\times\Bigg\{\sum_{\mathrm{pairs}\atop i,j} \sum_{k\neq i,j}
  {\cal D}_{ij,k}(p_a,p_b,p_1,...,p_{n+1})
  F_J^{(n)}(p_a,p_b,p_1,..,\tilde{p}_{ij},\tilde{p}_{k},..,p_{n+1})
  \Theta(y_{ij,k} < \alpha)\\ \nn
&&\quad\qquad+ \sum_{\mathrm{pairs}\atop i,j}
  \bigg[{\cal D}_{ij}^a(p_a,p_b,p_1,...,p_{n+1})
  F_J^{(n)}(\tilde{p}_a,p_b,p_1,..,\tilde{p}_{ij},..,p_{n+1}) 
  \Theta(1-x_{ij,a} < \alpha) \\ \nn
&& \hspace{2cm}
+ (a\leftrightarrow b)\bigg]\\ \nn
&&\quad\qquad+ \sum_{i\neq k}\left[
  {\cal D}_k^{ai}(p_a,p_b,p_1,...,p_{n+1})
  F_J^{(n)}(\tilde{p}_a,p_b,p_1,..,\tilde{p}_k,..,p_{n+1}) 
  \Theta(u_{i} < \alpha) + (a\leftrightarrow b)\right]\\ 
&&\quad\qquad+ \sum_{i}\left[
  {\cal D}^{ai,b}(p_a,p_b,p_1,...,p_{n+1})
  F_J^{(n)}(\tilde{p}_a,p_b,\tilde{p}_1,...,\tilde{p}_{n+1}) 
  \Theta(\tilde{v}_i < \alpha) + (a\leftrightarrow b)\right]\Bigg\}\;\;.
\nn \\  \label{dipole-terms}
\end{eqnarray}
The functions ${\cal D}_{ij,k}$, ${\cal D}_{ij}^a$, ${\cal D}_k^{ai}$
and ${\cal D}^{ai,b}$ are the dipole terms for the various
combinations for emitter and spectator.  $\sum_{ \{ n+1\}}$ denotes
the summation over all possible configurations for this
$(n+1)$-particle phase space which is labelled as $d\Gamma^{(n+1)}$
and the factor $S_{\{n+1\}}$ is the symmetry factor for identical
particles. In MadDipole, we have introduced four different $\alpha$-parameters, one
for each type of dipoles~\cite{Frederix:2008hu}. In our code  they are called
\texttt{alpha\_ff}, \texttt{alpha\_fi}, \texttt{alpha\_if} and
\texttt{alpha\_ii} for the final-final, finial-initial, initial-final
and initial-initial dipoles, respectively. The actual values for these
parameters are by default set to unity, corresponding to the original
formulation of the dipole subtraction
method~\cite{Catani:1996vz,Catani:2002hc}, but can be changed by the
user.

The integrated dipole factors, which are to be added with the virtual
$n$-parton contribution, also depend on $\alpha$. For case of massless
partons, the $\alpha$-dependence of the integrated terms is stated in
\cite{Nagy:1998bb,Nagy:2003tz} while for massive partons results for
most cases can be found in
\cite{Campbell:2004ch,Campbell:2005bb,Czakon:2009ss}. The remaining
cases, \textit{i.e.}, the (finite) massless-to-massive splittings, can
be found in the appendix.

\subsection{Regularization scheme dependence}
Calculating objects that contain divergences requires a systematic
prescription of how to deal with these divergences, \textit{i.e.}, a
scheme for their regularization. In NLO calculations, the same
regularization scheme has to be applied in the real emission part and
in the virtual corrections.  Both these contributions will differ
between different regularization schemes, while their sum
(\textit{i.e.}, the full NLO result) is scheme-independent.  Therefore
it is necessary to clearly specify which regularization scheme one is
using.

In QCD calculations, there are mainly two types of regularization
schemes used, namely dimensional regularization
\cite{'tHooft:1972fi,Bollini:1972ui,Ashmore:1972uj,Cicuta:1972jf} and
dimensional reduction
\cite{Siegel:1979wq,Siegel:1980qs,Stockinger:2005gx}. Both extend the
dimensionality of space-time to $d=4-2\e$, resulting in divergences
becoming explicit as poles in $1/\e$.  A discussion about their
subtypes and their differences can be found in \cite{Signer:2008va}.

In the real radiation contribution, the dependence on the
regularization scheme does not yet appear explicitly at the level of
the unintegrated dipole terms, and we consequently did not address
this issue in the previous release of
MadDipole~\cite{Frederix:2008hu}.

The helicity subroutines on which MadGraph and MadDipole are build
evaluate matrix elements in four dimensions. Therefore we can compute
the subtraction terms only in regularization schemes in which the
external particles are defined in four dimensions. The two most widely
used are the 't Hooft-Veltman scheme (tHV) in dimensional
regularization, which is the default used in our implementation, and
the four-dimensional helicity scheme (FDH). Both methods differ only
by a finite shift~\cite{Kunszt:1993sd,Catani:1996vz}:
\begin{equation}
{\cal V}_I(\ep)^{\scriptscriptstyle\rm tHV} \to {\cal V}_I^{\scriptscriptstyle\rm FDH}(\ep)
= {\cal V}_I(\ep)^{\scriptscriptstyle\rm tHV} - {\tilde \gamma}_I + {\cal O}(\ep) \;\;,
\end{equation}
\begin{equation}
{\tilde \gamma}_q = {\tilde \gamma}_{\bar q} = \frac{1}{2} \,C_F
\;, \;\;\;\;\;\;\; {\tilde \gamma}_g = \frac{1}{6} \,C_A \;\;.
\end{equation}
In the massive case there is no dependence on the regularization
scheme \cite{Catani:2000ef}. There is a simple flag in our code
that allows to change between these two schemes.

\section{Expansion of integrated dipoles in $\epsilon$}
\label{sec:exp}
Integration of the dipoles makes their infrared singularities explicit
as poles in the dimensional regularization parameter $\e$.  The formal
structure of the integrated dipoles is independent of the
configuration we are considering. For definiteness, we discuss only
the final-final case, the same structure also holds in all other
cases. For initial state hadrons, a additional collinear contribution
is present, which is rendered finite by mass factorization, which we
will describe here as well.

\subsection{Infrared final-state singularities}
In the final-final case, the integrated dipole function is written as
\begin{equation}
 \label{eq:Iijk_def}
\int [\rd p_i(\tpij,\tpk)] \,
\frac{1}{(p_i+p_j)^2-m_{ij}^2} \, \langle\bV_{ij,k}\rangle
\,\equiv\, \frac{\alps}{2\pi}\frac{1}{\Gamma(1-\eps)}
\biggl(\frac{4\pi\mu^2}{2\tpij\cdot \tpk}\biggr)^\eps {\cal V}_{ij}(\eps)\:.
\end{equation}
Depending on the configuration, the splitting function and the
propagator on the left hand side of (\ref{eq:Iijk_def}) change their
form; the structure of the result on the right hand side remains the
same.

The factor ${\cal V}_{ij}(\eps)$ is determined by the specific
configuration. It is singular in the limit $\epsilon \rightarrow 0$,
and is expanded as a Laurent series in $\epsilon$.  Prefactors taken out
from the expansion must be consistent between the dipoles and
the virtual one-loop corrections to the process under consideration.
 
In the $\overline{{\rm MS}}$-scheme, only universal factors are taken
out, and this expansion can be written symbolically as
\begin{flalign}
\int [\rd p_i(\tpij,\tpk)] &\,
\frac{1}{(p_i+p_j)^2-m_{ij}^2} \, \langle\bV_{ij,k}\rangle
\,\equiv\, \frac{\alps}{2\pi}\frac{1}{\Gamma(1-\eps)}
\biggl(\frac{4\pi\mu^2}{2\tpij \cdot \tpk}\biggr)^\eps {\cal V}_{ij}(\eps) \nn \\
&=
\frac{e^{\gamma\e}}{(4\pi)^\e}\,
\left(\frac{y_{ij;1}}{\epsilon^2}+\frac{y_{ij;2}}{\epsilon}+y_{ij;3}\right),
\label{eq:expy}
\end{flalign}
where $\gamma$ is the Euler-Mascheroni constant,
$\gamma=0.5772\ldots$.  This structure is the basis for our
implementation.

The specific values of $y_{ij;1}$, $y_{ij;2}$ and $y_{ij;3}$ depend on
the splitting one is considering.  For instance if one takes the
massless final-final quark-gluon splitting, \textit{i.e.}, where the
emitter is a massless quark int the final state, the unresolved
particle is a gluon, and the spectator is also a massless final state
particle, then the integrated splitting function is given by
\begin{equation}
\label{vi_ff}
 {\cal V}_{qg}(\eps) = \CF \left( \frac{1}{\eps^2}+\frac{3}{2\eps}+5-\frac{\pi^2}{2}\right).
\end{equation}
Consequently, the expansion coefficients in (\ref{eq:expy}) 
become
\begin{flalign}
 y_{qg;1}&=  \frac{\alps}{2\pi}\CF, \nn \\
 y_{qg;2}&= \frac{\alps}{2\pi}\CF\left(\frac{3}{2}+\log\left(\frac{\mu^2}{2\tpij \cdot \tpk}\right)\right), \nn \\
 y_{qg;3}&=\frac{\alps}{2\pi}\CF\left(\frac{1}{2}\log^2\left(\frac{\mu^2}{2\tpij \cdot \tpk}\right)+\frac{3}{2}
      \log\left(\frac{\mu^2}{2\tpij \cdot \tpk}\right)-\frac{7\pi^2}{12}+5\right)\;.
\end{flalign}

\subsection{Initial-state collinear behaviour}

The cases with initial state radiation, \textit{i.e.}, initial-initial
and initial-final, are slightly more involved. Not all singularities
that occur in the real emission process are cancelled by the virtual
corrections. The integrated initial-final dipole functions take the
form:
\begin{eqnarray}
\lefteqn{ \int [\rd p_i(\tpk;p_a,z)] \,
\frac{1}{(p_i+p_j)^2-m_{ij}^2} \, \frac{n_s(\widetilde{ai})}{n_s(a)}\, \langle\bV_{k}^{ai}\rangle}\nonumber \\
&\equiv& \frac{\alps}{2\pi}\frac{1}{\Gamma(1-\eps)}
\biggl(\frac{4\pi\mu^2}{2 p_a\cdot \tpk}\biggr)^\eps {\cal V}^{a,ai}(z;\eps)
\nonumber \\
&=&\frac{e^{\gamma\e}}{(4\pi)^\e}\,
\left(\frac{y_{i,j;1}(z)}{\epsilon^2}+\frac{y_{i,j;2}(z)}{\epsilon}+y_{i,j;3}(z)\right)\;\;.  
\end{eqnarray}
The left-over singularities arise from collinear splitting off the
initial emitter particle.  For example, for the initial-final dipole
describing the gluon radiation off an incoming quark, one has
\begin{eqnarray}
{\cal V}^{q,q}(z;\eps) & = &-\frac{1}{\e} P^{qq}(z) + \delta (1-z) \left[{\cal V}_{qg}(\e) 
+ \left(\frac{2\pi^2}{3} - 5 \right)\CF \right] + B^{q,q} (z) + {\cal O}(\e)\,,
\end{eqnarray}
where $B^{q,q} (z)$ contains regular functions and plus-distributions
in $z$.
 
The collinear singularity remains and is absorbed into the parton
distribution function. This is done by introducing a collinear
counterterm and its contribution to the cross section is given by
(6.6) of~\cite{Catani:1996vz}:
\begin{equation}
 \label{saC}
d\sigma_a^{C}(p;\mu_F^2) = - \frac{\as}{2 \pi} \;\frac{1}{\Gamma(1-\ep)} \sum_b
\int_0^1 dz \,
\left[ - \frac{1}{\ep}
\left( \frac{4 \pi \mu^2}{\mu_F^2} \right)^{\ep} P^{ab}(z) + \Kab(z) \right]
\,d\sigma_b^{B}(zp) \;,
\end{equation}
where in $\overline{{\rm MS}}$ scheme, $\Kab(z)=0$.

Neglecting the sum over the partonic subprocesses, we have a
counterterm contribution of the form
\begin{equation}
\label{ic}
 I^{ab}_c(\epsilon)=-\frac{\as}{2\pi}\frac{1}{\Gamma(1-\epsilon)}\left(\frac{4\pi\mu^2}{\mu_F^2}\right)^{\epsilon}
 \left[-\frac{1}{\epsilon}P^{ab}(z)\right]\;\;
\end{equation}
which is added to the integrated dipole.

In the $\overline{{\rm MS}}$ scheme the expansion of (\ref{ic}) is
given by
\begin{equation}
 I^{ab}_c(\epsilon)=\frac{e^{\gamma\e}}{(4\pi)^\e}\,
\left(\frac{y^c_{a,b;2}}{\epsilon}+y^c_{a,b;3}\right)\;.
\end{equation}
The coefficients $y^c_{a,b;2}$ and $y^c_{a,b;3}$ are given by
\begin{flalign}
 y^c_{a,b;2}&=\frac{\alpha_s}{2\pi}\cdot P^{ab}(z)\nn\\
 y^c_{a,b;3}&=\frac{\alpha_s}{2\pi}\cdot P^{ab}(z)\log\left(\frac{\mu^2}{\mu_F^2}\right)\;.
\end{flalign}

\subsection{Structure of program output}
The final output the program provided to the user is then given by all
contributions of the coefficients $y_{i}$ and $y^c_{i}$ respectively
multiplied with a born level matrix element that is modified by its
color structure. This explicitly means
\begin{quote}
\begin{tabular}{ll}
$\frac{1}{\epsilon^2}$:& \qquad $y_1\cdot\ _m\bra{1\cdots m}\frac{{\bom T}_i\cdot {\bom T}_k}{{\bom T}_i^2}\ket{1\cdots m}_m$\;, \\
$\frac{1}{\epsilon}$:& \qquad $(y_2+y^c_{2})\cdot\ _m\bra{1\cdots m}\frac{{\bom T}_i\cdot {\bom T}_k}{{\bom T}_i^2}\ket{1\cdots m}_m$\;,\\
finite: & \qquad $(y_3+y^c_{3})\cdot\ _m\bra{1\cdots m}\frac{{\bom T}_i\cdot {\bom T}_k}{{\bom T}_i^2}\ket{1\cdots m}_m$\;.\\
\end{tabular}
\end{quote} 

The contributions from the collinear counterterms are of course only
present if there are initial state QCD particles involved. By adding
the collinear counterterm, only endpoint singularities, which occur at
$z=1$, remain. Those then completely cancel with the virtual
corrections. The finite pieces can contain regular functions of $z$ as
well as $\delta$-functions and plus-distributions.

\section{Implementation and how to use the integrated dipoles}
\label{sec:prog}
The installation and running of the new package is very similar to the
already existing MadDipole package:

\begin{itemize}
\item[1.] Download the MadDipole package (version
  4.4.35 or later), \texttt{MG\_ME\_DIP\_V4.4.??.tar.gz}, from one of
  the MadGraph websites, {\it e.g.},
  \texttt{http://madgraph.phys.ucl.ac.be/}.
\item[2.] Extract and run \texttt{make} in the \texttt{MadGraphII}
  directory.
\item[3.] Copy the \texttt{Template} directory into a new directory,
  {\it e.g.}, \texttt{MyProcDir} to ensure that you always have a
  clean copy of the Template directory.
\item[4.] Go to the new \texttt{MyProcDir} directory and specify your
  process in the file \texttt{./Cards/proc\_card.dat}.  This is the
  $(n+1)$-particle process you require the subtraction term for.
\item[5.] Running \texttt{./bin/newprocess} generates the code for
  the $(n+1)$-particle matrix element and for all dipole terms and
  their integrated versions. After running this you will find a newly
  generated directory \texttt{./SubProcesses/P0\_yourprocess} ({\it
    e.g.}, \texttt{./SubProcesses/P0\_e+e-\_uuxg}) which contains all
  required files.
\end{itemize}

In the \texttt{./SubProcesses/P0\_yourprocess} directory all the files
relevant to that particular subprocesses are generated. In particular
this includes the $(n+1)$ particle matrix elements in the file
\texttt{matrix.f} and the dipoles in the files \texttt{dipol???.f},
where \texttt{???} stands for a number starting from
\texttt{001}. Furthermore the directory has two files,
\texttt{dipolsum.f} and \texttt{intdipoles.f}, where the sum of the
dipoles and their integrated versions are calculated, respectively.

Here we discuss in more detail the syntax and implementation of the
integrated dipoles in the \texttt{intdipoles.f} file. In this file
there are two subroutines of the form
\begin{quote}
 \texttt{intdipoles(P,\ X,\ Z,\ PSWGT\, EPSSQ,\ EPS,\ FIN)} \textrm{ and}\\
 \texttt{intdipolesfinite(P,\ X,\ Z,\ PSWGT\, EPSSQ,\ EPS,\ FIN)},
\end{quote}
where the input parameters are the phase space point
\texttt{P(0:3,nexternal)}, the Bjorken $x$ values of 
both incoming parton distributions, \texttt{X(1:2)}, and
the momentum fraction of the incoming parton that goes into the hard
process after an initial state radiation or if the spectator is an
initial state particle, \texttt{Z}.  It is the latter quantity which is
denoted with $x_{ij,a}$, $x_{ik,a}$ and $x_{i,ab}$ respectively
in~\cite{Catani:1996vz}, but we shall refer to it here simply as
$z$. In the numerical integration, it must be taken uniformly over the
interval $z\in [0;1]$ to ensure correct representation of the
distributions. Furthermore, the phase space weight should be passed as
well. These first four arguments should be provided by the user.
For the given phase space point, the routines evaluate the sum
of all integrated dipole subtraction terms (integrated dipole factors
multiplied with the appropriate reduced matrix elements) after mass
factorization of the collinear singularities. The routines call
external subroutines (explained below) from {\tt dipolesum.f} which
supply the parton distributions appearing with the reduced matrix
elements, apply event rejection cuts and pass the event information
into histograms.

The integrated dipoles in the two subroutines correspond to the
unintegrated dipoles in \texttt{dipolsum(..)}~and
\texttt{dipolsumfinite(..)}.  The dipoles in the
\texttt{dipolsumfinite(..)}~and
\texttt{intdipolesfinite(..)}~subroutines are not needed to cancel
singularities because they correspond to gluon splittings into massive
particles. However, they can be useful for checks when taking the
limit of vanishing quark masses, or to cancel some large logarithms in
the real emission matrix elements~\cite{Frederix:2008hu}.

The output parameters are the coefficients of the divergent and finite
terms: {\tt EPSSQ} is the coefficient of $1/\epsilon^2$, {\tt EPS} is
the coefficient of $1/\epsilon$. After inclusion of the collinear
counterterms they contribute with a factor $\delta(1-z)$, and are real
numbers. {\tt FIN} is the finite coefficient.  The calculation of the
\texttt{FIN} coefficient requires some explanation, since its coming
from distributions in $z$.  In the \texttt{intdipoles(..)} and
\texttt{intdipolesfinite(..)}  subroutines a three-dimensional array
\texttt{FINITE} is filled with the contributions from the various
dipoles:
\begin{quote}
{\tt FINITE(1)}: regular function in $z$,\\
{\tt FINITE(2)}: coefficient of $\delta(1-z)$,\\
{\tt FINITE(3)}: coefficient of $\delta(z_+-z)$.
\end{quote}
The last entry appears only for massive dipoles, with
\begin{equation}
 z_{+}=1-4\mu_Q^2,
\end{equation}
where $\mu_Q$ is the rescaled fermion mass occurring in the splitting.

In this decomposition, the $\delta$-functions and $(+)$-distributions
are carried out by taking into account the convolution with the
product of a reduced matrix element $g(z)$ (generated by MadDipole out
of MadGraph) and a parton distribution function $h(z)$ (supplied by
the user through the subroutine {\tt dipolepdf})
\begin{align}\label{split}
\int_0^1 {\rm d} z \, \delta(1-z) g(z) h(z) &= \int_0^1 {\rm d} z 
\underbrace{ g(1)\, h(1)}_{\text{in {\tt FINITE(2)}}}\,, \\
 \int_0^1 {\rm d} z \left(f(z)\right)_{+}g(z)\,h(z)&=
\int_0^1 {\rm d} z \Big[ \underbrace{f(z)\, g(z)\, h(z)}_{\text{ in {\tt FINITE(1)}}}
- \underbrace{f(z) \, g(1)\, h(1)}_{\text{in {\tt FINITE(2)}}}\Big].
\end{align}
In the massive case the point $z=1$ can not be reached in all cases
but we may have a reduced endpoint $z_+$, such that instead of a
$\delta(1-z)$ we then have a $\delta$-distribution of the form
$\delta(z_{+}-z)$ which is the third entry of the array of the finite
terms. If we have such a reduced endpoint then also the
$(+)$-distribution is generalized into a $z_{+}$-distribution as
defined in (\ref{eq:xpdist}). As before, we have the following
implementation:
\begin{align}
  \int_0^1 {\rm d} z \, \delta(1-z_{+}) g(z) h(z) &= \int_0^1 {\rm d}
  z \,
  \underbrace{ g(z_+)\, h(z_+)}_{\text{in {\tt FINITE(3)}}} \,,\\
  \int_0^1 {\rm d} z \left(f(z)\right)_{z_+}g(z)\,h(z)&= \int_0^1 {\rm
    d} z \,\,\Theta(z_+-z) \Big[ \underbrace{f(z)\, g(z)\,
    h(z)}_{\text{ in {\tt FINITE(1)}}} - \underbrace{f(z) \, g(z_+)\,
    h(z_+)}_{\text{in {\tt FINITE(3)}}}\Big].
\label{splitmassive}
\end{align}

By making a transformation of variables and computing the parton
distribution function not at $x$ (Bjorken's $x$) but at $x/z$, the
matrix element itself becomes independent of $z$ as the initial state
particle, that radiates the unresolved particle, then comes with the
momentum fraction of $x/z \cdot z=x$. Therefore the set of momenta
which are used to calculate the matrix element do not depend on
$z$. This means that $g(z)=g(z_+)=g(1)$ in
Eqs.~(\ref{split}--\ref{splitmassive}).

The final result \texttt{FIN} is the sum of the three contributions:
\begin{equation}
  \texttt{FIN}=\big(\texttt{FINITE(1)}/z+\texttt{FINITE(2)}+
  \texttt{FINITE(3)}/z_+\big)\times\texttt{PSWGT}.
\end{equation}
Since \texttt{FINITE(3)} appears only in massless-to-massive splittings, 
it is always zero for the \texttt{intdipoles(..)} subroutine.
Conversely,
the \texttt{EPSSQ} and \texttt{EPS} should be zero when coming from
the subroutine \texttt{intdipolesfinite(..)}.

For ease of use, we have provided three dummy subroutines/functions that the
user might want to fill when using the code. In the code there are
consistent calls to these subroutines. These subroutines/functions can be found
in the file \texttt{dipolsum.f} and are
\begin{description}
\item[\texttt{passcutsdip(P)}:] In this \texttt{LOGICAL FUNCTION} the
  user should provide a set of cuts that he wants to be applied to the
  phase space points. Note that in general the phase-space mapping for
  each unintegrated dipole is different; there needs to be a call to
  this function for each dipole. It should return \texttt{FALSE} if
  the point fails the cuts. By default every point passes the cuts.
\item[\texttt{dipolepdf(P,leg1,leg2,WGT)}:] In this \texttt{SUBROUTINE}
  the user should provide the value for his/her favourite PDF set for
  the two incoming particles with PDG codes passed by \texttt{leg1}
  and \texttt{leg2}. The factorization scale should be defined in the
  include file \texttt{dipole.inc}. The weight from the PDF should be
  returned in the argument \texttt{WGT}, which is set to \texttt{1} by
  default.
\item[\texttt{writehist(P,WGT)}:] In this \texttt{SUBROUTINE} the
  phase-space point (for each dipole) are provided together with its
  weight. The user could use these to fill histograms or save ntuples.
\end{description}
Besides these three subroutines/functions, the more technical
parameters, like the $\alpha$-parameter (see Sec.~\ref{sec:alpha}),
the number of flavors, the renormalization scheme and the scales can
be set in the include file \texttt{dipole.inc}. When changing any of
the parameters in this file the code should be recompiled (after
removing the object files) for these changes to take effect.

Besides the already existing \texttt{check} checking program,
that checks the limits of the real emission matrix element minus the
subtraction terms, we provide the user with another sample
program, \texttt{checkint}, to calculate
the value of the integrated subtraction terms for a given (or random)
phase space point.

More details and latest news, updates, bug fixes, etc.~can be found at
\begin{quote}
\texttt{http://cp3wks05.fynu.ucl.ac.be/twiki/bin/view/Software/MadDipole}.
\end{quote}

\section{Checks}
\label{sec:checks}
To verify the implementation we have performed two different kinds of
checks: independence on the phase space restriction parameter $\alpha$
and comparison with the implementation of dipoles in the MCFM
program~\cite{Campbell:1999ah,Campbell:2002tg}.

Both the non-integrated and the integrated dipoles depend on the
$\alpha$-parameter, the dependence on this parameter should cancel in
the sum. To validate this, we require:
\begin{equation}
\label{intalpha}
\int_{n+1} \left( d\sigma^R- d\sigma^A\right)+ \int_{n}
\left( \text{finite parts of int. dip.}\right)=\text{const},
\end{equation}
which must be a constant in that sense that it should not depend on
$\alpha$.  We have validated this for all 27 different cases
(emitter/spectator in initial/final state, and mass assignments)
listed in the appendix. Figure~\ref{fig:alpha} shows the dependence on
the $\alpha$-parameter for two examples: in the left plot is the
dependence shown on the $\alpha_{ii}$-parameter that governs the
initial-initial dipole phase-space restriction, and in the right plot
for a processes with massive spectators the dependence on the
$\alpha_{if}$-parameter (that restricts the initial-final dipole phase
space) is shown. The solid (red) lines display the quantity defined in
(\ref{intalpha}), which is observed to be independent on $\alpha$ over
several orders of magnitude.

\psfrag{alphaii}{$\alpha_{ii}$}
\psfrag{sigma}[][][1][-90]{$\sigma$}
\psfrag{TITLE}{\qquad \qquad
  \qquad $u\ \bar{u} \rightarrow e^{+}\ e^{-}\ g$ }
\psfrag{alphaif}{$\alpha_{if}$}
\psfrag{sigma1}[][][1][-90]{$\sigma$}
\psfrag{Graph}{\qquad \qquad
  \qquad $u\ \bar{u} \rightarrow t\ \bar{t}\ g$ }
\begin{figure}[t]
  \begin{center}
    \mbox{
      \subfigure[]{
        \epsfig{file=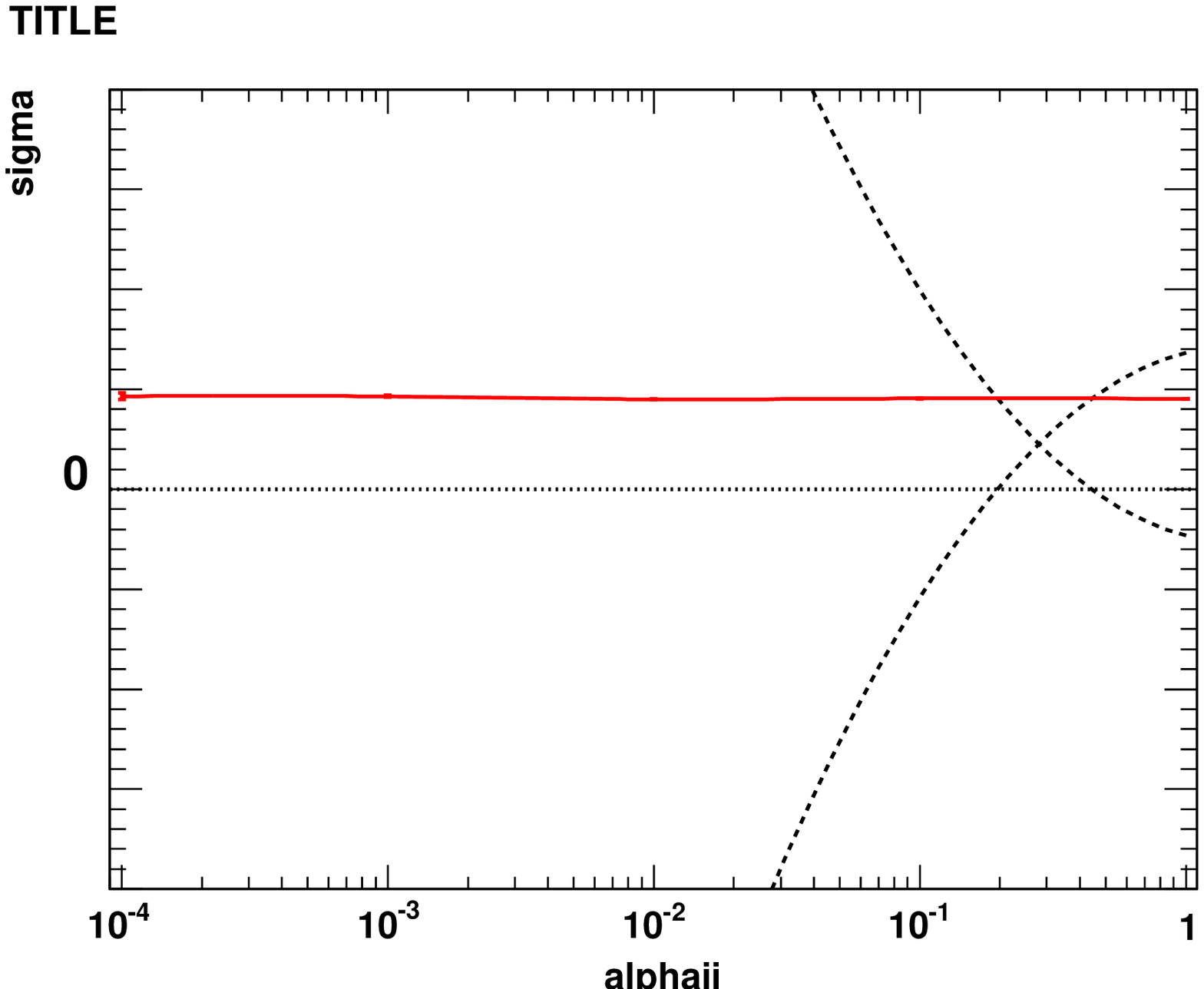, width=0.48\textwidth}
      }
      \subfigure[]{
        \epsfig{file=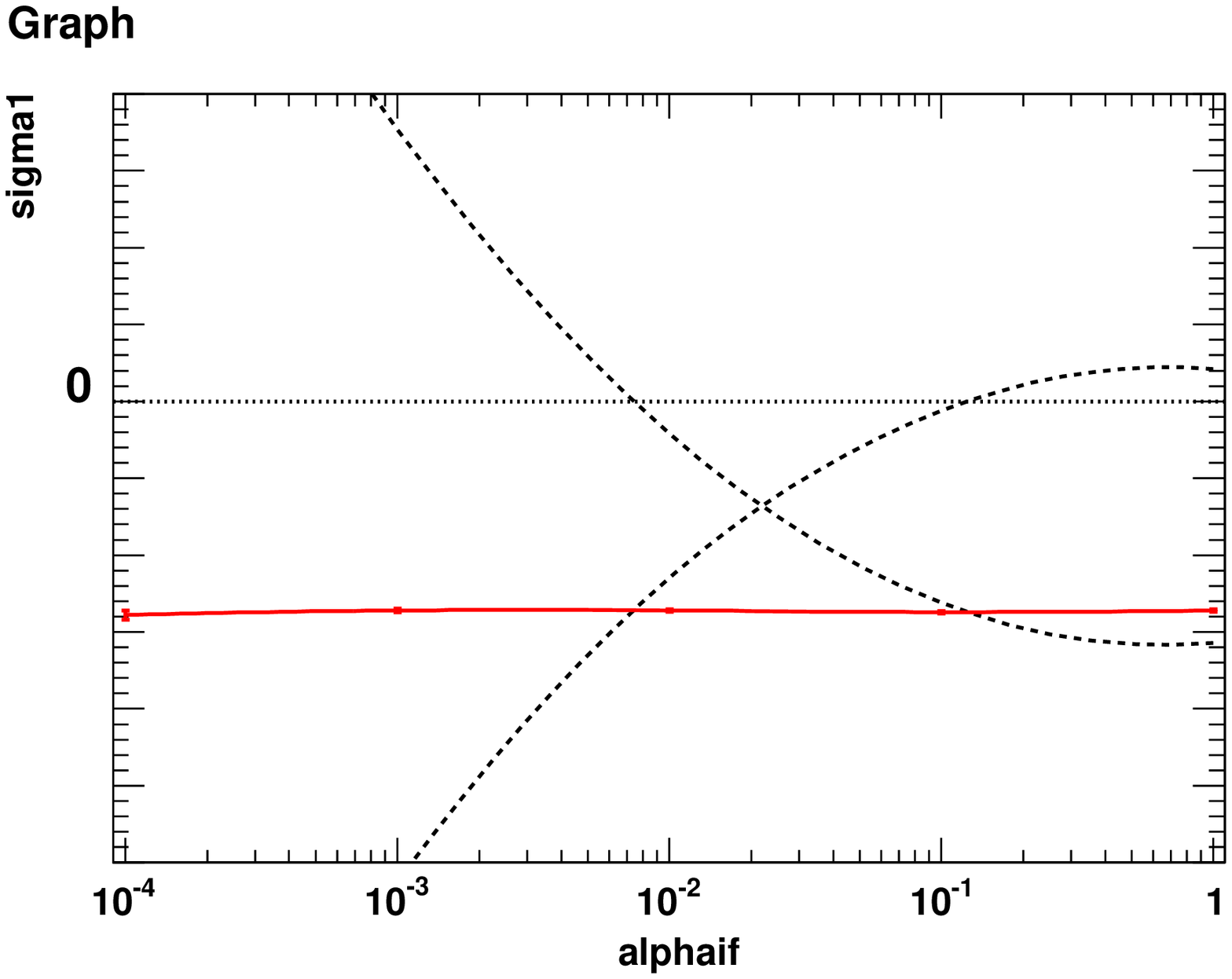, width=0.48\textwidth}
      }
    }
  \end{center}
  \vspace{-20pt}
  \caption{\label{fig:alpha} Two examples for the cancellation of the
    $\alpha$-dependence between unintegrated and integrated dipoles on
    specific contributions.  Since these do not correspond to full
    physical processes, we use an arbitrary normalization.  The plot
    on the left shows the $\alpha$ dependence of the initial-initial
    dipole where a gluon is radiated of the initial quark lines. On
    the right the $\alpha$ dependence of an initial-final dipole where
    the spectator is massive is shown. For both figures only the
    $\alpha$ parameter shown is varied, all others are kept fixed. The
    upper dashed line is the first contribution of
    (\protect\ref{intalpha}), \textit{i.e.}, the sum of matrix element
    and unintegrated dipoles. The lower dashed line is the second
    contribution, \textit{i.e.}, the finite terms of the integrated
    dipoles. The (red) solid line is the sum of both.
    For the sum we also included the Monte-Carlo error to show
    that the results for different values are consistent with each other. }
\end{figure}

This is a very powerful check because it includes several aspects of
the implementation. It verifies not only the correct implementation of
the $\theta$-functions for the unintegrated subtraction terms and the
$\alpha$-dependent correction terms for the integrated dipoles, but
independence on this parameter can only be achieved if the correct
parton distribution functions are called with the right arguments, and
if the various terms contributing to the integrated dipoles are summed
correctly. Moreover, an inconsistent infrared-unsafe implementation of
the cuts (by the user) will lead to a dependence on the $\alpha$
parameter.

The $\alpha$-independence does, however, not probe important features
related to the pole structure of the subtraction terms, namely the
scheme dependence and the factorization scale dependence, neither does
it show other finite contributions. Therefore our second check was a
direct comparison of our implementation against MCFM
\cite{Campbell:1999ah,Campbell:2002tg}.  We have compared the results
for single phase space points, which allowed us to directly probe the
implementation of the integrated terms.  The various terms (regular
functions, $\delta$- and plus-distributions) could be checked
separately.  We note that not all possibilities (massless/massive,
initial/final emitter/spectator) are present in MCFM and therefore not
all could be checked.  This includes in particular the finite dipoles:
massless-to-massive splittings have no divergences, but do have a
potentionally large logarithm which might be useful to subtract.

\section{Conclusions}
\label{sec:conc}

The MadDipole package~\cite{Frederix:2008hu} provides dipole
subtraction terms for the evaluation of real radiation corrections in
NLO calculations within the framework of the MadGraph/MadEvent matrix
element and event
generator~\cite{Stelzer:1994ta,Maltoni:2002qb,Alwall:2007st}. In this
work, we described the extension of the MadDipole package to include
the integrated dipole terms, which are required to combine the real
radiation corrections to a given process with the virtual corrections
and collinear counterterms of the parton distributions. With the newly
developed subroutines, MadDipole provides the unintegrated and
integrated dipole subtraction terms.

The integrated subtraction terms are convoluted with the user-supplied
parton distributions and are combined with the collinear counterterms
from mass factorization.  Consequently, infrared singularities appear
only in the kinematic endpoints, which correspond to the kinematics of
the virtual corrections. The MadDipole output can thus be readily
combined with the results from one-loop matrix element generators,
which are currently under rapid
development~\cite{Ossola:2007ax,Ossola:2008xq,Berger:2008sj,Giele:2008bc,Binoth:2008uq}.

A first application of the MadDipole package, in combination with the
GOLEM one-loop amplitude generator~\cite{Binoth:2008uq}, was in the
calculation of NLO corrections to the quark-antiquark contribution to
$pp\to b\bar b b\bar b$~\cite{Binoth:2009rv}. Many more applications
are likely to follow.

\section*{Acknowledgements}
We would like to thank Fabio Maltoni for many useful discussions and John
Campbell for some help on the checks against MCFM. NG wants to thank
the University of Zurich for the kind hospitality during final work on
the project.  This research was supported in part by the Swiss
National Science Foundation (SNF) under contract 200020-126691 and in
part by the U.~S.~Department of Energy under contract
No.~DE-FG02-91ER40677.

\begin{appendix}
\section{$\alpha$-dependence of the integrated dipoles}
In this appendix we give a list of the references,
 where the $\alpha$-dependent terms can be found in the
literature. For the few cases which were not known previously, we give the
result below.

\subsection{Final-final}
\begin{quote}
\begin{tabular}{cc@{ }lll}
\toprule
\multicolumn{3}{c}{Configuration} & Reference & Remarks \\
\midrule
(a) &$Q \to Q\ g$,& $m_k>0$& \cite{Bevilacqua:2009zn} (A.9,A.17,A.20) & Different definition of $\alpha$\\
(b) &$Q \to Q\ g$,& $m_k=0$ & \cite{Campbell:2004ch} (A.22,A.23) & \\
(c) &$q \to q\ g$,& $m_k>0$ & \cite{Campbell:2004ch} (A.29,A.31) & \\
(d) &$q \to q\ g$,& $m_k=0$ & \cite{Nagy:1998bb} (22) & \\
(e) &$g \to g\ g$,& $m_k>0$ & \cite{Bevilacqua:2009zn} (A.9,A.19,A.22) & Different definition of $\alpha$\\
(f) &$g \to g\ g$,& $m_k=0$ & \cite{Nagy:1998bb} (24) & \\
(g) &$g \to Q\ \bar{Q}$,& $m_k>0$ & this work, (\ref{ffgQQmasive}) & \\
(h) &$g \to Q\ \bar{Q}$,& $m_k=0$ & this work, (\ref{ffgQQmassless}) & \\
(i) &$g \to q\ \bar{q}$,& $m_k>0$ & \cite{Bevilacqua:2009zn} (A.18,A.21) & Different definition of $\alpha$\\
(j) &$g \to q\ \bar{q}$,& $m_k=0$ &  \cite{Nagy:1998bb} (23) & \\
\bottomrule
\end{tabular}
\end{quote}
In the massive case there is an ambiguity how one defines
$\alpha$. The upper limit of the integration over $y$ is given by the
variable $y_{+}$.  One can now define $\alpha$ in such a way that the
maximal value for $\alpha$ is given by $y_{+}$. In that case the
integration range is given by the interval $y \in
[\alpha,y_{+}]$. Another possibility is to have $\alpha=1$ as the
maximal value which implies that the allowed integration range is
given by
\begin{equation}
\label{yint}
 \alpha y_{+}\ <\ y_{ij,k}\ <\ y_{+}.
\end{equation}
The first definition has been used in \cite{Bevilacqua:2009zn} whereas
we use the latter.

The two cases (g) and (h) are in principle not needed as the result is
still finite due to the masses of the quarks.  For the sake of
completeness we also included this cases and therefore give the
result.

\subsubsection{Case (g)}
Case (g) is the splitting of a gluon into two massive quarks ($g
\rightarrow Q\ \bar{Q}$) with a massive spectator. The splitting
function for that process is given in Eq.(5.18)
of~\cite{Catani:2002hc},
\begin{flalign}
\label{gQQ}
\langle\bV_{Q\bar Q,k}\rangle  =
8\pi\mu^{2\eps}\alps\TR& \frac{1}{\vijk}
\Biggl\{ 1-\frac{2}{1-\eps}
\Biggl[ \zi(1-\zi)-(1-\kappa)z_+z_-
\nn\\& \qquad {}
-\frac{\kappa\mu_Q^2}{2\mu_Q^2+(1-2\mu_Q^2-\mu_k^2)\yijk} \Biggr]
\Biggr\}\:,
\end{flalign}
where we neglect the last term because of $\kappa=0$ in our
implementation. The integration ranges are
\begin{flalign}
 \frac{2\mu_j^2}{1-2\mu_j^2-\mu_k^2}\ <\ &y_{ij,k}\ <\ y_{+} = 1-\frac{2\mu_k(1-\mu_k)}{1-2\mu_j^2-\mu_k^2},\\
 \frac{1-v_{ij,i}v_{ij,k}}{2}\ <\ &\tilde{z_i}\ < \frac{1+v_{ij,i}v_{ij,k}}{2}.
\end{flalign}
Here, the integration over $y_{ij,k}$ does not start at $y_{ij,k}=0$
but at a value larger than zero. But introducing the $\alpha$
parameter and calculating the correction terms implies new integration
boundaries for the $y_{ij,k}$ integration according to (\ref{yint}):
\begin{equation}
 \frac{2\mu_j^2}{1-2\mu_j^2-\mu_k^2}\ <\ \alpha y_{+},
\end{equation}
which implies that $\alpha$ must not be chosen to be too small for
this splitting.  With this restriction, we find \footnotesize
\begin{flalign}
\label{ffgQQmasive}
 I^{({\rm g})}_{ij,k}(\epsilon,\alpha)&=I^{({\rm g})}_{ij,k}(\epsilon)-\TR\left(\left(b d \left(-8 a \mu_j^4 
   \log \left(\alpha  c^2 y_{+}-d \sqrt{c^2-4 \mu_j^4}-4 \mu_j^4\right)+2 a c^2 \log
   \left(\alpha  c^2 y_{+}-d \sqrt{c^2-4 \mu_j^4}-4 \mu_j^4\right)
\right. \right. \right. \nn \\ & \left. \left. \left.
+2 a c \log \left(\alpha  c^2 y_{+}-d
   \sqrt{c^2-4 \mu_j^4}-4 \mu_j^4\right)+4 a \mu_j^2 \log \left(\alpha  c^2 y_{+}-d \sqrt{c^2-4
   \mu_j^4}-4 \mu_j^4\right)
\right. \right. \right. \nn \\ & \left. \left. \left.
-2 a \left(c^2+c-4 \mu_j^4+2 \mu_j^2\right) \log \left((1-\alpha  y_{+})
   \left(-c-2 \mu_j^2\right)^{5/2} \left(c-2 \mu_j^2+1\right)\right)
\right. \right. \right. \nn \\ & \left. \left. \left.
+8 a \mu_j^4 \log \left(-b \sqrt{c^2-4
   \mu_j^4}+c^2 y_{+}-4 \mu_j^4\right)-2 a c^2 \log \left(-b \sqrt{c^2-4 \mu_j^4}+c^2 y_{+}-4
   \mu_j^4\right)
\right. \right. \right. \nn \\ & \left. \left. \left.
-2 a c \log \left(-b \sqrt{c^2-4 \mu_j^4}+c^2 y_{+}-4 \mu_j^4\right)-4 a \mu_j^2 \log
   \left(-b \sqrt{c^2-4 \mu_j^4}+c^2 y_{+}-4 \mu_j^4\right)
\right. \right. \right. \nn \\ & \left. \left. \left.
+2 a \left(c^2+c-4 \mu_j^4+2 \mu_j^2\right)
   \log \left((y_{+}-1) \left(-\left(-c-2 \mu_j^2\right)^{5/2}\right) \left(c-2 \mu_j^2+1\right)\right)
\right. \right. \right. \nn \\ & \left. \left. \left.
-3 c^2
   \sqrt{2 \mu_j^2-c} \log (-2 (\alpha  c y_{+}+d))+4 c \mu_j^2 \sqrt{2 \mu_j^2-c} \log (-2 (\alpha  c
   y_{+}+d))
\right. \right. \right. \nn \\ & \left. \left. \left.
-2 c \sqrt{2 \mu_j^2-c} \log (-2 (\alpha  c y_{+}+d))+3 c^2 \sqrt{2 \mu_j^2-c} \log (-2 (b+c
   y_{+}))
\right. \right. \right. \nn \\ & \left. \left. \left.
-4 c \mu_j^2 \sqrt{2 \mu_j^2-c} \log (-2 (b+c y_{+}))+2 c \sqrt{2 \mu_j^2-c} \log (-2 (b+c
   y_{+}))\right)
 \right. \right. \nn \\ & \left. \left. 
+2 b d \sqrt{2 \mu_j^2-c} \left(c^2-2 (c+1) \mu_j^2+4 \mu_j^4\right) \tan ^{-1}\left(\frac{2
   \mu_j^2}{\sqrt{\alpha ^2 c^2 y_{+}^2-4 \mu_j^4}}\right)
 \right. \right. \nn \\ & \left. \left. 
+c \sqrt{2 \mu_j^2-c} \left(c^2 y_{+}
   \left(\alpha ^2 b y_{+}-2 \alpha  b-d (y_{+}-2)\right)+4 c \mu_j^2 (b (\alpha  y_{+}-1)+d (-y_{+})+d)+4
   \mu_j^4 (b-d)\right)
 \right. \right. \nn \\ & \left. \left. 
-2 b d \sqrt{2 \mu_j^2-c} \left(c^2-2 (c+1) \mu_j^2+4 \mu_j^4\right) \tan
   ^{-1}\left(\frac{2 \mu_j^2}{\sqrt{c^2 y_{+}^2-4 \mu_j^4}}\right)\right)
 \right. \nn \\ & \left. 
/\left(3 c \left(2 \mu_j^2-c\right)^{3/2}
   \sqrt{c^2 y_{+}^2-4 \mu_j^4} \sqrt{\alpha ^2 c^2 y_{+}^2-4 \mu_j^4}\right)\right)
\end{flalign}
\normalsize
where we used the following abbreviations:
\begin{flalign*}
a&=\sqrt{1-\mu_k^2}\\
b&=\sqrt{c^2y_{+}^2-4\mu_j^4}\\
c&=-1+2\mu_j^2+\mu_k^2\\
d&=\sqrt{\alpha^2c^2y_{+}^2-4\mu_j^4}\;.
\end{flalign*}

\subsubsection{Case (h)}
Case (h) describes also the splitting of a gluon into a massive quark
pair ($g\rightarrow Q\ \bar{Q}$) however with a massless
spectator. While the splitting function is the same (\ref{gQQ}), the
integration boundaries are different, namely
\begin{flalign}
 \frac{2\mu_i^2}{1-2\mu_i^2}\ <\ &y_{ij,k}\ < 1 \\
 \frac{1-v_{ij,i}}{2}\ <\ &\tilde{z_i}\ <\ \frac{1+v_{ij,i}}{2}.
\end{flalign}
The phase space integral gets multiplied by $\Theta(\alpha-y_{ij,k})$
and integration leads to \footnotesize
\begin{flalign}
\label{ffgQQmassless}
 I^{({\rm h})}_{ij,k}(\epsilon,\alpha)&=I^{({\rm h})}_{ij,k}(\epsilon)-\TR\left(\frac{2}{3} \left(\frac{2 \sqrt{\alpha ^2 
\left(1-2 \mu_j^2\right)^2-4 \mu_j^4}}{2 (\alpha -1)\mu_j^2-\alpha }
+\sqrt{\alpha ^2 \left(1-2 \mu_j^2\right)^2-4 \mu_j^4}+\left(2
   \mu_j^2-1\right)
\right.\right. \nn \\ & \left. \left.
 \left(-\log \left(-2 \left(\sqrt{\alpha ^2 \left(1-2 \mu_j^2\right)^2-4
   \mu_j^4}+\alpha  \left(2 \mu_j^2-1\right)\right)\right)+2 \tan ^{-1}\left(\frac{2
   \mu_j^2}{\sqrt{\alpha ^2 \left(1-2 \mu_j^2\right)^2-4 \mu_j^4}}\right)
\right.\right.\right. \nn \\ & \left. \left.\left.
+\log \left(-2
   \left(2 \mu_j^2+\sqrt{1-4 \mu_j^2}-1\right)\right)-2 \tan ^{-1}\left(\frac{2
   \mu_j^2}{\sqrt{1-4 \mu_j^2}}\right)\right)+\sqrt{1-4 \mu_j^2}\right)\right).
\end{flalign}
\normalsize

\subsection{Final-initial}
\begin{quote}
\begin{tabular}{c@{ }r@{ $\to$ }lll}
\toprule
\multicolumn{3}{c}{Configuration} & Reference & Remarks\\
\midrule
(a)& $Q$&$Q\ g$&  \cite{Campbell:2004ch} (A.13,A.14) & \\
(b)& $q$&$q\ g$&  \cite{Campbell:2004ch}(A.17) / \cite{Nagy:2003tz} (11-16)& Different approaches\\
(c)& $g$&$Q\ \bar{Q}$& this work, (\ref{eq:I_QQa},\ref{eq:JfiQQacont},\ref{JQQ_nonsing_a})  & \\
(d)& $g$&$q\ \bar{q}$&  \cite{Nagy:2003tz} (11-16) & Different approach in MadDipole\\
(e)& $g$&$g\ g$&  \cite{Nagy:2003tz} (11-16)& Different approach in MadDipole\\
\bottomrule
\end{tabular}
\end{quote}
The cases (b), (d), and (e) can be found in \cite{Nagy:2003tz}.  Their
result however contains already the sum of different contributions
making use of the $\boldmath{I}$\mdseries - and
$\boldmath{K}$\mdseries -flavor kernels. As it mixes different
contributions this is not suitable for the MadDipole implementation.
Therefore we use \cite{Campbell:2004ch} for (b) and derive results for
(d) and (e) following the approach in \cite{Campbell:2004ch}.  Of
course, both approaches are equivalent.  Again, for the sake of
completeness we also add the finite case (c).
 
\subsubsection{Case (c)}
The one particle phase space for final-initial dipoles is given in
Eq.(5.48) of~\cite{Catani:2002hc}:
\begin{flalign}
 \int [\rd p_i(\tpij;p_a,x)] &=
\frac{1}{4} (2\pi)^{-3+2\eps} 
(2\tpij p_a)^{1-\eps}
\int_0^{x_+} \rd \xija\, \delta(x-\xija) \,
(1-x+\mu_{ij}^2)^{-\eps} 
\nn\\ & \times \int\rd^{d-3}\Omega\int_{z_-(x)}^{z_+(x)}\rd\zi\, 
\left[z_+(x)-\zi\right]^{-\eps}\left[\zi-z_-(x)\right]^{-\eps},
\label{eq:psfi}
\end{flalign}
and the integrated splitting function is given by Eq.(5.53)
of~\cite{Catani:2002hc} as:
\begin{equation}
\label{eq:Iija_def}
\int [\rd p_i(\tpij;p_a,x)] \,
\frac{1}{(p_i+p_j)^2-m_{ij}^2} \, \langle\bV_{ij}^a\rangle
\equiv \frac{\alps}{2\pi}\frac{1}{\Gamma(1-\eps)}
\biggl(\frac{4\pi\mu^2}{2\tpij p_a}\biggr)^\eps 
I_{ij}^a(x;\eps) \;.
\end{equation}
For case (c), the splitting of a gluon into massive quarks
($g\rightarrow Q\ \bar{Q}$), the integrated spitting function is given
in Eq.(5.57) of~\cite{Catani:2002hc}:
\begin{equation}
 \label{eq:I_QQa}
I_{Q\bar Q}^a(x;\ep) = \TR \left\{
[J_{Q\bar Q}^a(x, \mu_Q)]_{x_+} + \delta(x_+-x)\,
\Big[J_{Q\bar Q}^{a;\rm S}(\mu_Q;\ep) + J_{Q\bar Q}^{a;\rm NS}(\mu_Q)\Big]
\right\}
+\Oe{}\:,
\end{equation}
where the $x_{+}$-distribution is defined as:
\begin{equation}
 \label{eq:xpdist}
\int_0^1\rd x\, \Big(f(x)\Big)_{x_+} g(x) \equiv
\int_0^1\rd x\, f(x) \Theta(x_+ - x) \left[g(x)-g(x_+)\right]\:.
\end{equation}
Imposing the cut on the $\alpha$-parameter implies that the phase
space in (\ref{eq:psfi}) is multiplied with
$\Theta(\alpha-1+x_{ija})$.

This leads to a modification of the $x_{+}$-distribution terms and we
get in analogy to Eq.(5.62) of~\cite{Catani:2002hc}
\begin{equation}
 [J_{Q\bar Q}^a(x, \mu_Q,\alpha)]_{x_+} = 
\frac{2}{3} 
\left(
\frac{1-x+2\mu_Q^2}{(1-x)^2} 
\sqrt{1-\frac{4\mu_Q^2}{1-x}}
\right)_{1-\alpha}\:,
\label{eq:JfiQQacont}
\end{equation}
where we define:
\begin{equation}
 \int\limits_0^1dx\ f(x)\left(g(x)\right)_{1-\alpha}=\int\limits_{1-\alpha}^1dx\ g(x)\left(f(x)-f(1)\right)
\end{equation}
The non-singular terms $J_{Q\bar Q}^{a;\rm NS}(\mu_Q)$ receive:
\footnotesize
\begin{flalign}
\label{JQQ_nonsing_a}
 J_{Q\bar Q}^{a;\rm NS}(\mu_Q,\alpha)&=J_{Q\bar Q}^{a;\rm NS}(\mu_Q)+
\frac{2}{9} \left(\left(-4 \mu_Q ^2 \left(\sqrt{\frac{\left(1-4
   \mu_Q ^2\right) \left(\alpha -4 \mu_Q ^2\right)}{\alpha
   ^3}}+4\right)-5 \sqrt{\frac{\left(1-4 \mu_Q ^2\right)
   \left(\alpha -4 \mu_Q ^2\right)}{\alpha }}
\right.\right.\nn \\ & \left.\left.
-16 \mu_Q
   ^4+5\right)/\left(\sqrt{1-4 \mu_Q ^2}\right)
+6 \log \left(\sqrt{\alpha -4 \mu_Q
   ^2}+\sqrt{\alpha }\right)-6 \log \left(\sqrt{1-4 \mu_Q
   ^2}+1\right)\right).
\end{flalign}
\normalsize

\subsubsection{Case (d)}
Case (d) is just the limit $\mu_Q\rightarrow 0$ of
(\ref{eq:JfiQQacont}), \textit{i.e.},
\begin{equation}
 [J_{q\bar q}^a(x, 0,\alpha)]_{+} = 
\frac{2}{3} \left(\frac{1}{1-x}\right)_{1-\alpha},
\end{equation}
which leads to the following additional non-singular terms:
\begin{equation}
 J_{q\bar q}^{a;\rm NS}(0,\alpha)=J_{q\bar q}^{a;\rm NS}(0)+\frac{2}{3}\log \alpha.
\end{equation}

\subsubsection{Case (e)}
In the case of the splitting ($g\rightarrow g\ g$) the general structure of the
integrated splitting function is  given by Eq.(5.66) of~\cite{Catani:2002hc}:
\begin{equation}
 \label{eq:I_gga}
I_{gg}^a(x;\ep) = 2\CA \left\{
[J_{gg}^a(x)]_{+} + \delta(1-x)\,J_{gg}^{a;\rm S}(\ep)
\right\}+\Oe{} \:.
\end{equation}
The first term $[J_{gg}^a(x)]_{+}$ contains all $+$-distributions but is
not a $+$-distribution itself. In the presence of the $\alpha$-parameter we
find
\begin{equation}
 \label{eq:Jfiggacont}
[J_{gg}^a(x,\alpha)]_+ = 
\left(\frac{2}{1-x}\ln\frac{1}{1-x}-\frac{11}{6}\frac{1}{1-x} \right)_{1-\alpha}
+\frac{2}{1-x}\ln(2-x)\Theta(\alpha-1+x)\:,
\end{equation}
which leads to a modification of the terms proportional to
$\delta(1-x)$ of Eq.(5.68) of~\cite{Catani:2002hc} of the following
form:
\begin{equation}
 J_{gg}^{a;\rm S}(x,\alpha)=J_{gg}^{a;\rm S}(x)-\log^2\alpha-\frac{11}{6}\log\alpha.
\end{equation}

\subsection{Initial-final}
\begin{quote}
\begin{tabular}{clll}
\toprule
\multicolumn{2}{c}{Configuration} & Reference & Remarks\\
\midrule
(a)& $\tilde{ij}:\ q,\quad \text{emitter}:\ q$ &  \cite{Campbell:2004ch} (A.9, A.11)/ \cite{Nagy:2003tz} (11-16) & \cite{Nagy:2003tz} only massless\\ 
(b)& $\tilde{ij}:\ g,\quad \text{emitter}:\ q$ &  \cite{Campbell:2005bb} (A.8)& \\
(c)& $\tilde{ij}:\ q,\quad \text{emitter}:\ g$ &  \cite{Campbell:2005bb} (A.10)& \\
(d)& $\tilde{ij}:\ g,\quad \text{emitter}:\ g$ &  \cite{Campbell:2005bb} (A.11)& \\
\bottomrule
\end{tabular}
\end{quote}
From the analytical point of view the limit of a vanishing spectator
mass can be performed without any problems. However taking the result
for a massive spectator and setting the mass to zero in the numerical
implementation causes problems for the cases (b) and (d). For these
two cases we calculated the limit analytically and implemented a
massive and a massless version.

\subsection{Initial-initial}
\begin{quote}
\begin{tabular}{clll}
\toprule
\multicolumn{2}{c}{Configuration} & Reference & Remarks\\
\midrule
(a)& $\tilde{ij}:\ q,\quad \text{emitter}:\ q$ &  \cite{Campbell:2004ch} (A.4)/ \cite{Nagy:2003tz} (11-16) & \\ 
(b)& $\tilde{ij}:\ g,\quad \text{emitter}:\ q$ &  \cite{Campbell:2004ch} (A.5)/ \cite{Nagy:2003tz} (11-16)& \\
(c)& $\tilde{ij}:\ q,\quad \text{emitter}:\ g$ &  \cite{Campbell:2005bb} (A.4)/ \cite{Nagy:2003tz} (11-16)& \\
(d)& $\tilde{ij}:\ g,\quad \text{emitter}:\ g$ &  \cite{Campbell:2005bb} (A.5)/ \cite{Nagy:2003tz} (11-16)& \\
\bottomrule
\end{tabular}
\end{quote} 
For our implementation the results from \cite{Campbell:2004ch,Campbell:2005bb} are used.

\end{appendix}

\bibliographystyle{utphys}
\bibliography{references}
\end{document}